\documentclass[conference]{IEEEtran}
\usepackage{amsmath}
\usepackage{amssymb}
\usepackage{graphicx}
\usepackage{epstopdf}
\usepackage[tight]{subfigure}
\graphicspath{{figures/}}

\begin{document}
\title{Flip-OFDM for Optical Wireless Communications}
\date{}

\author{\IEEEauthorblockN{Nirmal Fernando}
\IEEEauthorblockA{Dept. Elec. and Comp. Sys.\\
Monash University\\
Clayton, VIC 3800\\
Email: Nirmal.Fernando@Monash.edu}
\and
\IEEEauthorblockN{Yi Hong}
\IEEEauthorblockA{Dept. Elec. and Comp. Sys.\\
Monash University\\
Clayton, VIC 3800\\
Email: Yi.Hong@Monash.edu}
\and
\IEEEauthorblockN{Emanuele Viterbo}
\IEEEauthorblockA{Dept. Elec. and Comp. Sys.\\
Monash University\\
Clayton, VIC 3800\\
Email: Emanuele.Viterbo@Monash.edu}}

\IEEEspecialpapernotice{(Invited Paper)}

\maketitle

\begin{abstract}
We consider two uniploar OFDM techniques for optical wireless communications: asymmetric clipped optical OFDM (ACO-OFDM) and Flip-OFDM. Both techniques can be used to compensate multipath distortion effects in optical wireless channels. However, ACO-OFDM has been widely studied in the literature, while the performance of Flip-OFDM has never been investigated. In this paper, we conduct the performance analysis of Flip-OFDM and propose additional modification to the original scheme in order to compare the performance of both techniques. Finally, it is shown by simulation that both techniques have the same performance but different hardware complexities. In particular, for slow fading channels, Flip-OFDM offers 50\% saving in hardware complexity over ACO-OFDM at the receiver.
\end{abstract}


\section{Introduction}

Orthogonal Frequency Division Multiplexing (OFDM) has been widely studied as a technology to compensate dispersion effects in the optical wireless communication \cite{armstrong_ofdm_2009}. In optical wireless communication, intensity modulation with direct detection (IM/DD) technique is commonly used for data transmission. However, IM/DD communication is non-coherent (i.e. phase of the optical carrier can not be used to transmit information) and transmit signal must be {\em real} and {\em positive}. These additional constraints require some special care, if OFDM is to be used in optical wireless communications, since the equivalent baseband time-domain OFDM signal is usually  complex.

The most common technique to generate a real time-domain signal is to preserve the Hermitian symmetry property of the OFDM subcarriers at the expense of loosing half of the available bandwidth. Although this technique produces a real time domain signal, this signal is bipolar and needs to be converted into unipolar signal before the transmission. The traditional method of converting the bipolar OFDM signal into an unipolar symbol is to add a DC bias. This is known as DC-offset OFDM (DCO-OFDM) \cite{armstrong_ofdm_2009}. However, the amplitude of the required DC bias depends on the peak to average power ratio (PAPR) of the OFDM symbol, and since OFDM has a high PAPR ratio, the amount of DC bias is generally significant. As shown in \cite{armstrong_comparison_2008}, the large DC-bias makes DCO-OFDM optical power inefficient. On the other hand, the use of lower DC bias makes large negative time samples clipped, which may result in considerable inter-carrier interference and out of band optical power.

Asymmetric Clipped Optical OFDM (ACO-OFDM) proposed in \cite{armstrong_power_2006} can avoid any DC bias requirement. In ACO-OFDM, only odd subcarriers carry information symbols and any negative values are clipped at the transmitter. It is shown in \cite{armstrong_power_2006} that clipping the time domain signal does not distort symbols in odd subcarriers, although their amplitude is scaled by a half. In \cite{armstrong_performance_2006, xia_li_channel_2007, armstrong_comparison_2008}, the performance of ACO-OFDM is compared to other modulation schemes such as on-off keying and DC-biased OFDM (DC-OFDM). The results suggest that ACO-OFDM has better power efficiency than any other modulation scheme for optical wireless channels \cite{armstrong_comparison_2008}. Performance of ACO-OFDM can be further improved by using bit loading schemes, as presented in \cite{wilson_digital_2008, wilson_transmitter_2009}.

It can be seen that research community has followed ACO-OFDM as the main unipolar OFDM technique for optical wireless communication. An alternative unipolar OFDM technique proposed in \cite{yong_modulation_2007} has been largely ignored to the best of our knowledge. We will name this technique as Flip-OFDM. In Flip-OFDM, the positive and negative parts are extracted from the bipolar OFDM real time-domain signal, and transmitted in two consecutive OFDM symbols. Since the negative part is flipped before transmission, both subframes have positive samples. Flip-OFDM is thus a unipolar OFDM technique that can be used in optical wireless communications.

This paper provides three main contributions: (i) we review and analyze Flip-OFDM; (ii) we suggest further improvements to Flip-OFDM; (iii) we finally compare the performance and complexity of  Flip- and ACO-OFDMs.


\section{System model}
\subsection{Optical Wireless Channel}
\begin{figure}[t!]
        \centering%
        \includegraphics[scale=.53]{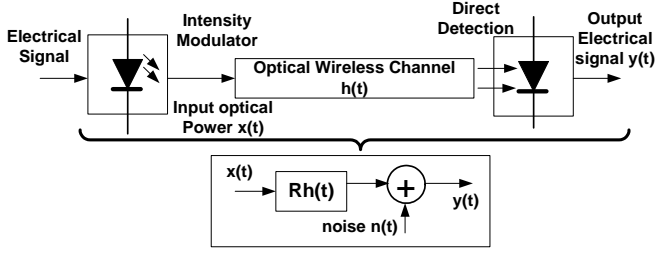}
        \caption{Equivalent based band channel model for IM/DD optical wireless channels}
        \label{IR_BB_channel_model}
      \end{figure}

A block diagram of a typical IM/DD based optical wireless communication system is shown in Fig. \ref{IR_BB_channel_model}. In most optical wireless applications, an infrared emitter is used as optical transmitter to generate optical signal $x(t)$. This signal represents the intensity of the optical carrier transmitted over optical wireless channel. At the receiver, a photodetector collects the optical signal and converts it to an electrical current $y(t)$.

In general, the optical wireless link can be operated in two modes: directed and non directed (or diffused) \cite{kahn_wireless_1997}. In directed optical wireless link, the contribution of the Line-of-Sight (LOS) is dominating and an additive white Gaussian noise (AWGN) channel model is appropriate \cite{jungnickel_physical_2002}. However, the AWGN assumption is no longer valid for diffused optical wireless channels, where no strong LOS is present. The optical wireless channel is modeled as a linear baseband system \cite{kahn_wireless_1997, carruthers_modeling_1997}, as shown in Fig. \ref{IR_BB_channel_model}. Given the channel impulse response and noise component $h(t)$ and $n(t)$, respectively, the received electrical signal can be given as,
      \begin{equation}
            \label{base_band_channel_model}
                   y(t) = x(t) \ast h(t) + n(t)
      \end{equation}
where $\ast$ denotes convolution.

\subsubsection{Diffused Optical Wireless Channel Model}
In this case, the propagation of optical power along various paths of different lengths contributes to the multipath distortion. As stated in \cite{carruthers_modeling_1997}, using series of Dirac delta functions, the impulse response of a optical wireless channel can be modeled as,
        \begin{equation}
            \label{OW_channel_response}
            h(t) = \sum_{n=0}^{N_D}{h_n \delta(t- n \Delta\tau)}\\
        \end{equation}
where $h_n$, $N_D$ and $N_D \times \Delta\tau$ represent channel coefficients, number of paths and the maximum delay of the channel, respectively. There are two physical processes that produce multipath dispersion: i) the light captured through multiple reflections and ii) the light captured from single reflection.

Alternatively, the exponential-decay model and ceiling-bounce model proposed in \cite{carruthers_modeling_1997} are widely used to model both two multiple and single optical reflections.
In the exponential-decay model, the channel impulse response due to multiple reflections is modeled as
        \begin{equation}
           \label{exponential_decay_model}
            h_e(t)  = \frac{1}{D}e^{-\frac{t}{D}}u(t)
        \end{equation}
where $D$ and $u(t)$ represent the rms delay spread of the multiple reflections and unit step function, respectively. Furthermore, in ceiling-bounce model, the channel impulse response is given by
        \begin{equation}
            \label{ceiling_bounce_model}
             h_c(t,a) = \frac{6a^6}{(t+a)^7}u(t)
        \end{equation}
where $a = 12 \sqrt{\frac{11}{13}} D$.

\subsubsection{Noise Model}
In optical wireless communication systems, the two dominant noise components are photon noise and receiver circuit thermal noise \cite{kahn_wireless_1997, carruthers_modeling_1997}. The photon noise is due to the discreteness of photon arrivals, which is mainly due to background light sources. Although the received background light can be minimized through optical filtering, even in a well designed photo detector, it creates shot noise. Since the background light power is usually more powerful than the transmitted signal, the contribution of the transmitted optical signal to this shot noise is negligible \cite{kahn_wireless_1997, carruthers_modeling_1997}. Therefore, it is common to assume that the shot noise is independent of the transmitted signal and is modeled as white Gaussian. On the other hand, when there is no background light, the dominant noise source is the thermal noise from the receiver preamplifier and it is in both signal-independent and Gaussian. Thus, in both cases, the total noise can be well modeled as Gaussian and signal-independent. The power spectral density of the shot noise and thermal noise are given in \cite{kahn_wireless_1997}.

\subsection{Flip-OFDM}
\begin{figure}[t!]
      \centering%
      \includegraphics[scale=.43]{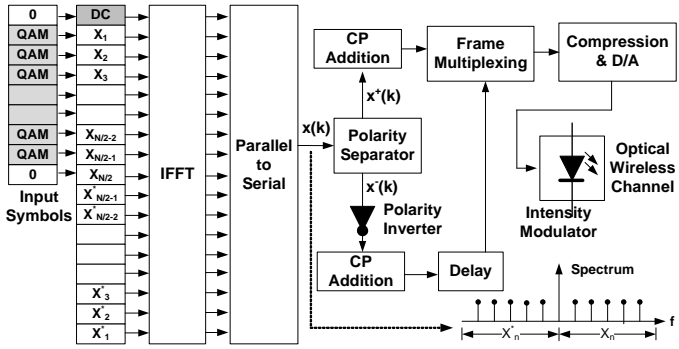}
      \caption{Block Diagram of Flip-OFDM Transmitter}
      \label{flipped_tx}
   \end{figure}

\begin{figure}[htb!]
      \centering%
      \includegraphics[scale=.42]{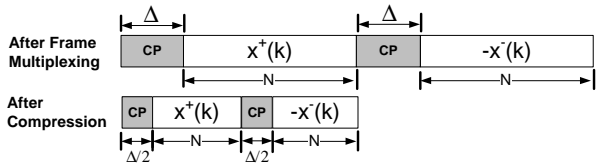}
      \caption{Flip-OFDM Unipolar Frame}
      \label{flipped_ofdm_frame}
   \end{figure}

\begin{figure}[t!]
      \centering%
      \includegraphics[scale=.43]{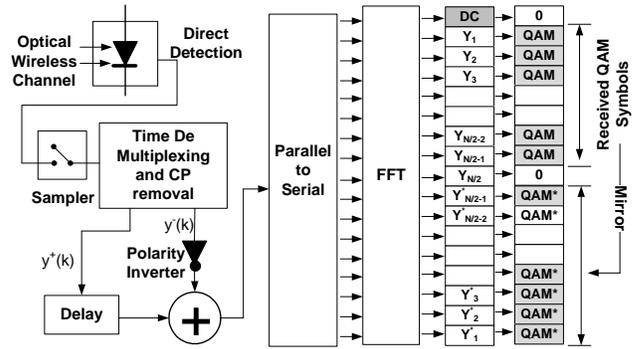}
      \caption{Block Diagram of Flip-OFDM Receiver}
      \label{flipped_rx}
     \end{figure}

Fig. \ref{flipped_tx} shows a block diagram of a Flip-OFDM transmitter. Let $X_n$ be the transmitted QAM symbol in the $n$-th OFDM subcarrier. The output of Inverse Fast Fourier Transform (IFFT) operation at the $k$-th time instant is given by
    \begin{equation}
        \label{genearl_ofdm_equation}
          x(k)  = \sum_{n=0}^{N-1}{X_n\exp{(\frac{j 2 \pi n k}{N})}}
    \end{equation}
where $N$ is the IFFT size. If the symbols $X_n$ contained in each OFDM subcarrier are independent, the time domain signal $x(k)$ produced by the IFFT operation is complex. This can be avoided by imposing the Hermitian symmetry property, i.e,
    \begin{equation}
        \label{hermitian_symmetry_lbl}
           X_{n} = X^\ast_{N-n} \hspace{4mm} n = 0,1,2 ... , N/2-1
    \end{equation}
where $^{\ast}$ denotes complex conjugation. Hence, half of OFDM subcarriers have to be sacrificed to generate the real time-domain signal. The output of IFFT operation in (\ref{genearl_ofdm_equation}) is then given as,
     \begin{eqnarray}
       \label{HS_ofdm_equation}
       x(k)  & = & X_0 + \sum_{n=1}^{N/2-1}{X_n\exp{(\frac{j 2 \pi n k}{N})}} + X_{N/2} \exp{(j \pi k)} \nonumber  \\
             & + & \sum_{n=N/2+1}^{N-1}{{X^*_{N-n}}\exp{(\frac{j 2 \pi n k}{N})}}
     \end{eqnarray}
where $X^*_n$, $X_0$ and $X_{N/2}$ represent the conjugate symmetric QAM symbol of $X_n$, the DC component and the QAM symbol of center carrier, respectively. In addition, to avoid any DC shift or any residual complex component in the time domain signal, $X_0$ and $X_{N/2}$ in (\ref{HS_ofdm_equation}) are set to zero.
Flip-OFDM uses half of the total OFDM subcarriers to carry information so that
the output of IFFT block is a real bipolar signal $x(k)=x^+(k) + x^-(k)$, where the positive and negative parts are defined as,
    \begin{eqnarray}
     \nonumber x^+(k) &=& \begin{cases}
             x(k) & \text{if } x(k) \geq 0 \\
             0 &  \text{otherwise}\\
     \end{cases} \\
     x^-(k) &=& \begin{cases}
             x(k) & \text{if }x(k) < 0 \\
             0 &  \text{otherwise}\\
     \end{cases}
     \end{eqnarray}
and $k = 1,2,... ,N$.
The positive signal  $x^+(k)$ is transmitted in the first OFDM subframe, while the second OFDM subframe is used to carry the flipped  (inverted polarity) signal $-x^-(k)$,  as shown in Fig. \ref{flipped_tx}. Since the communication happens over a dispersive optical channel, cyclic prefixes of duration  $\Delta$ are added to both OFDM subframes. The second OFDM subframe is then delayed by $(N+\Delta)$ and time multiplexed after the first one.
All the samples in both subframes are unipolar OFDM symbols. In the original proposal of \cite{yong_modulation_2007}, the time samples after frame multiplexing are compressed in time by a factor two, to produce the same symbol duration of a single bipolar OFDM frame length (Fig. \ref{flipped_ofdm_frame}). In the next section we will discuss the implication of such operation. We simply note here that the compressed guard interval should accommodate the full delay spread of the channel.
At the Flip-OFDM receiver, the two received subframes are used to reconstruct the bipolar OFDM frame as shown in Fig. \ref{flipped_rx}. The cyclic prefixes associated with each OFDM subframe are first removed and the original bipolar signal is regenerated as,
     \begin{equation}
        \label{regen_bipolar_frm}
        y(k) = y^+(k) - y^-(k)
     \end{equation}
where $y^+(k)$ and $y^-(k)$ represent the respective time samples belonging to first and second subframes. Fast Fourier Transform (FFT) operations are performed to recreate the bipolar signal in order detect the transmitted information symbols at the receiver.

\section{Comparison of ACO- and Flip-OFDM}

To make a fair comparison, we first modify Flip-OFDM \cite{yong_modulation_2007} so that both ACO- and Flip-OFDMs have the same bandwidth, data rate, and the cyclic prefix length. We then compare some key parameters of both systems: {\em i)} spectral efficiency,  {\em ii)} electrical domain signal-to-noise-ratio (SNR) and
{\em iii)} the expected Bit Error Rate (BER) performance. We assume that both systems are used on the same optical wireless channel with a given delay spread and that the channel is constant over two consecutive OFDM symbols.

\subsection{Modification of Flip-OFDM}
The Flip-OFDM in \cite{yong_modulation_2007} performs compression of time samples, so that it matches to the original unipolar symbol duration, as shown in Fig. \ref{flipped_ofdm_frame}. Although the compression doubles the bandwidth and data rate, it reduces the length of the cyclic prefix by half when compared to ACO-OFDM.

\begin{figure}[t!]
      \centering%
      \includegraphics[scale=.45]{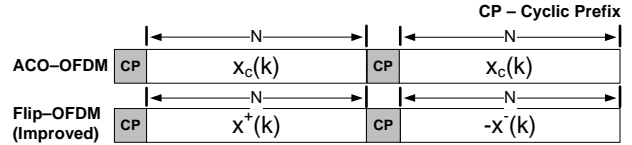}
      \caption{OFDM frame structure of ACO-OFDM and Flip-OFDM. We assume FFT and IFFT sizes of both ACO- and Flip-OFDM are the same. $N$ denotes the FFT and IFFT size for each case.}
      \label{ofdm_frame_struct}
\end{figure}

Different from \cite{yong_modulation_2007}, we do not compress the two OFDM frames, as shown in Fig. \ref{ofdm_frame_struct}. Hence, the cyclic prefixes are the same for both systems. Moreover, the bandwidth of both systems is the same, given the same FFT/IFFT sizes and sampling rates.

\subsection{Comparison of key parameters}

\subsubsection{Spectral Efficiency ($\eta$)}
Spectral efficiency is defined as the number of information bits per unit bandwidth and measured in [bits/Hz]. Both ACO- and Flip-OFDM sacrifice half of the spectrum in order to have a real bipolar time domain signal due to the Hermitian symmetry (\ref{hermitian_symmetry_lbl}). In ACO-OFDM, only odd subcarriers are used to transmit information and only positive time samples are sufficient to extract the information. Hence, the spectral efficiency of ACO-OFDM is one fourth of what could be achieved in a typical OFDM system with complex signals, i.e., ($1/4 \times \log_2(M)$). On the other hand, although Flip-OFDM uses both odd and even subcarriers, it needs two subframes to reconstruct the bipolar signal and to extract the information. Hence, Flip-OFDM uses  twice the number of samples of ACO-OFDM to transmit twice as many information symbols. Therefore, we can conclude that the spectral efficiencies of both ACO- and Flip-OFDMs are the same.

\subsubsection{Electrical domain SNR and expected BER performance}
The expected BER performance is directly related to electrical domain SNR, defined as ${E[{x_t}^2(k)]}/{\sigma^2}$, where $E[{x_t}^2(k)]$ gives the energy of the transmitted signal $x_t (k)$ and $\sigma^2$ represents the variance of the electronic noise. We can see that this electrical domain SNR depends on two parameters: i) equivalent electrical energy per symbol and ii) noise power, as discussed below.
\begin{itemize}
  \item
As a result of the asymmetric clipping in ACO-OFDM, half of the transmitted energy is shifted to even subcarriers as clipping noise \cite{armstrong_power_2006, armstrong_performance_2006}. Hence, each OFDM subcarrier has an average equivalent electrical energy of $E[{x_t}^2(k)]$, and only the in odd subcarriers carry useful information. Thus half of the transmitted energy appears to be wasted.
On the other hand, in Flip-OFDM, the equivalent electrical energy per symbol spread across the two OFDM symbols. At the receiver, this spread electrical energy is recombined during the regeneration of bipolar OFDM symbol. Therefore, the equivalent electrical energy per symbol is $2 \times E[{x_t}^2(k)]$, and is twice of that in ACO-OFDM. However, as shown later, the noise power is also doubled during the regeneration of the bipolar OFDM symbol and thereby both schemes have the same electrical domain SNR.
  \item
Unlike ACO-OFDM, although transmitted energy is not wasted in Flip-OFDM, the noise power is doubled during the recombination of positive and negative components.
In Flip-OFDM, assuming that $H^+_n$ and $H^-_n$ represent channel responses of $n$-th OFDM subcarrier over two subframes, the outputs of the $n$-th OFDM subcarrier in the two subframes are,
      \begin{eqnarray}
        \label{flip_rx_positive}
                  Y^+_n  &=& H^+_n X^+_n + Z_n^+\\
        \label{flip_rx_negative}
                  Y^-_n  &=& -H^-_n X^-_n + Z_n^-
      \end{eqnarray}
where $Z_n^+$ and $Z_n^-$ represent the noise components of $n$-th OFDM subcarrier. If we assume that the channel is constant over two consecutive OFDM symbols (i.e. $H^+_n = H^-_n \triangleq H_n$), then the addition of (\ref{flip_rx_positive}) and (\ref{flip_rx_negative}) gives,
     \begin{eqnarray}
        \label{noise_double_eqn_const_channel}
             R_n  &=& H_n X_n + \{ Z_n^+ + Z_n^-\}
      \end{eqnarray}
and standard channel equalization can be used to detect original symbols. As seen in (\ref{noise_double_eqn_const_channel}), the noise power of received symbol has been doubled even though transmitted symbols energy is not wasted.
\end{itemize}

In summary, we can see that both ACO- and Flip-OFDM shall have the same BER performance in the electrical domain. Specifically, in ACO-OFDM, as a result of the asymmetric clipping, half of transmitted energy is shifted to even subcarriers. On the other hand, in Flip-OFDM, although transmitted energy is fully usable, the noise power is doubled during the regeneration of bipolar OFDM symbol at the receiver.

\section{Simulation Results}
In this section, we compare the BER performance and hardware computation complexity of both ACO- and Flip-OFDMs.

      \begin{table}
      \begin{center}
      \caption{Simulation parameters}
      \begin{tabular}{l|c|c}
        \hline	
        Parameter &  \multicolumn{2}{c}{value} \\
        \cline{2-3}
                                                    & Diffused     & LOS (AWGN)\\
        \hline
        \hline
        $N$                                         & 256          & 256\\
        Symbol constellation mapping                & QPSK         & QPSK\\
        Sampling time ($T_s$)                       & .75 ns       & .75 ns\\
        Cyclic prefix($\Delta$)                     & $(64+1)T_s $ & $10T_s $\\
        $\Delta\tau$                                & .75 ns       & -\\
        Max. delay spread ($N_D \times \Delta\tau$) & 48ns         & -\\
        RMS delay spread ($D$)                       & 8 ns         & -\\
        \hline
      \end{tabular}
      \label{Simulation_parameters}
      \end{center}
      \end{table}

\subsection{BER}
\begin{figure}[htb!]
          \centering%
          \includegraphics[scale=.62]{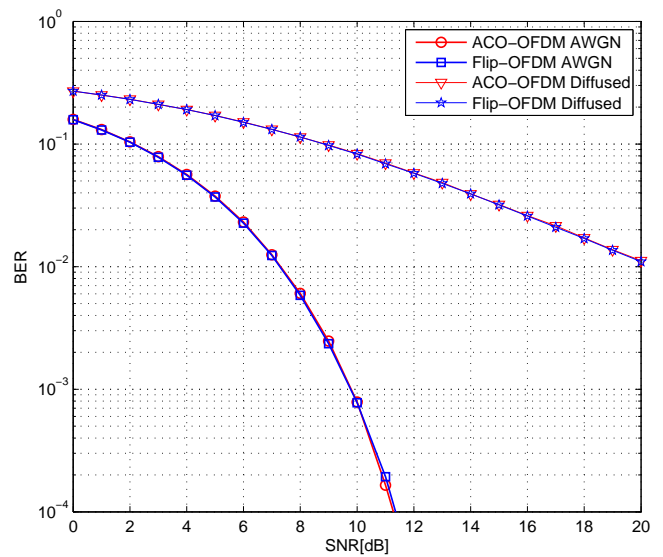}
          \caption{Bit error rate comparison of ACO and Flip-OFDM. Half of the channel coefficients in (\ref{OW_channel_response}) generated with Ceiling-Bounce Model and other half with Exponential-Decay Model. $h_n$ in (\ref{OW_channel_response}) is uniformly distributed from $0$ to $h_e(t)$ or $h_c(t,a)$. Key simulation parameters are given in Table \ref{Simulation_parameters}}
          \label{BER_perfomance_compare}
          \end{figure}

In Fig. \ref{BER_perfomance_compare}, we show the BER performance of ACO-OFDM and Flip-OFDM for Direct AWGN and Diffused optical wireless channels. The summary of key simulation parameters are given in Table \ref{Simulation_parameters}. The optical wireless channel is normalized so that  $\|h(t)\|^2 = 1$ in diffused mode of operation. It can be seen that electrical BER performance are the same for ACO-OFDM and Flip-OFDM in both AWGN and diffused wireless optical channels.

\subsection{Complexity of TX and RX operations}
We define complexity as the number of FFT/IFFT operations in the transmitter or the receiver. Table \ref{Hardware_complexities} provides a comparison of the hardware computation complexities in the transmitter and the receiver sites. At the transmitter, we see that both schemes have the same hardware computation complexities, if the IFFT operation in the ACO-OFDM is optimized by the fact that half of the subcarriers are set to zero. However, at the receiver, Flip-OFDM has a 50\% of hardware computation savings compared to ACO-OFDM.
    \addtocounter{footnote}{1}
    \begin{table}
    \begin{center}
    \caption{Hardware computation comparisons}
    \begin{minipage}{\linewidth} \center
    \begin{tabular}{l|l|l}
    \hline
                          & ACO-OFDM & Flip-OFDM \\
    \hline \hline

             Transmitter & $N\log(N)$         & $N\log(N)$ \\
             Receiver    & $2N\log(N)$        & $N\log(N)$ \\
    \hline
    \end{tabular}
    \end{minipage}
    \label{Hardware_complexities}
    \end{center}
    \end{table}

\section{Conclusion}

We analyzed a unipolar OFDM technique, named Flip-OFDM, which has been largely ignored in the open literature. We modified the Flip-OFDM and made comparisons with ACO-OFDM in terms of Bit Error Rate (BER) performances, spectral efficiencies and hardware computation complexities. We showed that both schemes have the same spectral efficiencies and BER performance in electrical domain but different hardware computation complexities. In particular, for slow fading, Flip-OFDM has a saving of 50\% in receiver complexity over ACO-OFDM.


\begin{thebibliography}{1}

\bibitem{armstrong_ofdm_2009}
J.~Armstrong, ``{OFDM} for optical communications,'' \emph{J. Lightwave
  Technol.}, vol.~27, no.~3, pp. 189--204, Feb. 2009.

\bibitem{armstrong_comparison_2008}
J.~Armstrong and B.~Schmidt, ``Comparison of asymmetrically clipped optical
  {OFDM} and {DC-Biased} optical {OFDM} in {AWGN},'' \emph{Communications
  Letters, {IEEE}}, vol.~12, no.~5, pp. 343--345, 2008.

\bibitem{armstrong_power_2006}
J.~Armstrong and A.~Lowery, ``Power efficient optical {OFDM},''
  \emph{Electronics Letters}, vol.~42, no.~6, pp. 370--372, 2006.

\bibitem{armstrong_performance_2006}
J.~Armstrong, B.~Schmidt, D.~Kalra, H.~Suraweera, and A.~Lowery, ``Performance
  of asymmetrically clipped optical {OFDM} in {AWGN} for an intensity modulated
  direct detection system,'' in \emph{Global Telecommunications Conference,
  2006. {GLOBECOM} '06. {IEEE}}, 2006, pp. 1--5.

\bibitem{xia_li_channel_2007}
X.~Li, R.~Mardling, and J.~Armstrong, ``Channel capacity of {IM/DD} optical
  communication systems and of {ACO-OFDM},'' in \emph{Communications, 2007.
  {ICC} '07. {IEEE} International Conference on}, 2007, pp. 2128--2133.

\bibitem{wilson_digital_2008}
S.~Wilson and J.~Armstrong, ``Digital modulation techniques for optical
  {Asymmetrically-Clipped} {OFDM},'' in \emph{Wireless Communications and
  Networking Conference, 2008. {WCNC} 2008. {IEEE}}, 2008, pp. 538--542.

\bibitem{wilson_transmitter_2009}
------, ``Transmitter and receiver methods for improving asymmetrically-clipped
  optical {OFDM},'' \emph{Wireless Communications, {IEEE} Transactions on},
  vol.~8, no.~9, pp. 4561--4567, 2009.

\bibitem{yong_modulation_2007}
J.~Yong, ``Modulation and demodulation apparatuses and methods for wired /
  wireless communication,'' Korea Patent WO2007/064\,165 A, 07, 2007.

\bibitem{kahn_wireless_1997}
J.~Kahn and J.~Barry, ``Wireless infrared communications,'' \emph{Proceedings
  of the {IEEE}}, vol.~85, no.~2, pp. 265--298, 1997.

\bibitem{jungnickel_physical_2002}
V.~Jungnickel, V.~Pohl, S.~Nonnig, and C.~von Helmolt, ``A physical model of
  the wireless infrared communication channel,'' \emph{Selected Areas in
  Communications, {IEEE} Journal on}, vol.~20, no.~3, pp. 631--640, 2002.

\bibitem{carruthers_modeling_1997}
J.~Carruthers and J.~Kahn, ``Modeling of nondirected wireless infrared
  channels,'' \emph{Communications, {IEEE} Transactions on}, vol.~45, no.~10,
  pp. 1260--1268, 1997.
  
\end{thebibliography}


\end{document}